# Kinetic theory of Hyaluronan cleavage by Bovine Testicular Hyaluronidase in Standard and Crowded Environments

Reine Nehmé[1], Rouba Nasreddine[1], Lucija Orlic [1,2], Chrystel Lopin-Bon[1], Josef Hamacek[2]*, Francesco Piazza[2]*

[1] Institut de Chimie Organique et Analytique (ICOA), CNRS FR 2708 - UMR 7311, Université d'Orléans, Orléans, France.

[2] Centre de Biophysique Moléculaire (CBM), CNRS UPR 4301, Université d'Orléans, 45071 Orléans, France

* Corresponding authors

    Josef.Hamacek@cnrs-orleans.fr

    Francesco.Piazza@cnrs-orleans.fr

*Abbreviations.*

**BTH,** Bovine testes hyaluronidase (hyaluronoglucosaminidase, EC 3.2.1.35); **CPA,** Corrected Peak Area; **CE,** Capillary electrophoresis ; **PEG**, polyethylenglycol MW 6000 Da ; **HA**, hyaluronic acid; **dimer** = disaccharide; **tetramer** = tetrasaccharide ; **hexamer** = hexasaccharide ; **octamer** = octasaccharide ; **decamer** = decasaccharide ; **oligomer** = oligosaccharide.

**Structured abstract**

**Background**

Details of the kinetic pathways governing enzymatic cleavage of hyaluronic acid (HA) by hyaluronidase are still widely uncharted. Capillary electrophoresis-based assays were used for accurate quantification of enzymatic products. A crowding agent was also employed to mimic excluded-volume constraints typical of in-vivo conditions.

**Scope**

Introduce a comprehensive kinetic model describing the late-stage degradation of HA by hyaluronidase and identify the relevant kinetic pathways and the associated rates.

**Major Conclusions**

All relevant fragmentation and transglycosylation pathways and rates were identified. Two dimers forming a tetramer is the dominant recombination pathway. Macromolecular and self-crowding slow down the kinetics but do not alter the underlying mechanisms.

**General Significance**

Our results bring a novel and comprehensive quantitative insight into enzymatic HA degradation. Rationalizing the effect of crowding brings the intricate conditions of in-vivo settings a little closer, and also stands as a powerful tool to pinpoint relevant kinetic pathways in complex systems.

**Abstract**

In this paper, we introduce a comprehensive kinetic model describing the enzymatic cleavage of hyaluronan (HA) by bovine testicular hyaluronidase (BTH). Our theory focuses specifically on the late stage of the hydrolysis, where the concentrations of a limited number of oligomers may be determined experimentally with accuracy as functions of time.

The present model was applied to fit different experimental sets of kinetic data collected by capillary electrophoresis at two HA concentrations and three concentrations of PEG crowder (0, 10, 17 % w/w). Our theory seems to apply universally, irrespective of HA concentration and crowding conditions, reproducing to an excellent extent the time evolution of the individual molar fractions of oligomers. Remarkably, we found that the reaction mechanism in the late degradation stage essentially reduces to the cleavage or transfer of active dimers. While the recombination of dimers is the fastest reaction, the rate-limiting step turns out to be invariably the hydrolysis of hexamers. Crowding, HA itself or other inert, volume-excluding agents, clearly boosts recombination events and concomitantly slows down all fragmentation pathways.

Overall, our results bring a novel and comprehensive quantitative insight into the complex reaction mechanism underlying enzymatic HA degradation. Importantly, rationalizing the effect of crowding not only brings the intricate conditions of *in-vivo* settings a little closer, but also emerges as a powerful tool to help pinpointing relevant kinetic pathways in complex systems.

# 1. Introduction

The enzymatic activity of hyaluronidases (Hyals) has been the object of studies performed under different conditions and using various analytical methods [1-3]. The overall reaction mechanism is complex, since the enzyme may catalyse both substrate hydrolysis as well as transglycosylation steps. Hyaluronic acid or hyaluronan (HA) is a polymer composed of repeating disaccharide units (D-glucuronic acid and N-acetyl-D-glucosamine, here denoted as dimers). The degradation of HA as a starting compound produces a high number of transient intermediates that are finally transformed into tetrasaccharide and/or disaccharide products, depending on the enzyme nature. The overall kinetics of HA degradation can be accessed by monitoring the total concentration of oligosaccharides possessing N-acetyl-D-glucosamine reducing ends as a function of time, since one enzymatic cleavage of O-glycosidic bonds produces one new reducing end. Several spectrophotometric assays have been developed for this purpose [1, 4, 5]. However, the use of separation techniques such as chromatography or capillary electrophoresis (CE) enables one to get extremely accurate insight into the degradation process by determining the molecular mass average [6] or directly the concentrations of intermediates at different reaction times [7]. While the measurements gathered from both these approaches can be used for evaluating the kinetic constants of the enzyme ($V_{max}$, $K_M$), the determination of HA-oligomer concentrations during the reaction allows one to obtain a deeper understanding of the reaction mechanism.

Several researchers have investigated the degradation of short, well-defined oligomers bearing the same motif as HA [8, 9]. The separation of reaction products catalyzed by Hyal revealed the transfer of the key disaccharide moiety between different substrates/products before converging to the tetrasaccharide end-product. While the mechanism for relatively short substrates is now qualitatively understood, access to the individual rate constants related to successive cleavage/recombination steps is still limited. The kinetics of Hyals strongly depends on the concentration of substrates and on the length of the substrate chains forming the complex with the enzyme [10, 11]. According to the present understanding, an initial phase of chaotic, fast hydrolysis of relatively high $M_r$ substrates [8] is followed by slower reactions involving shorter-length substrate chains, which is related to a lower stability of the enzyme-substrate complex. The maximum initial rate of degradation was found at $M_r$ (HA) ~ 20-200 kDa (~100-100 monomers) [11]. On the other hand, transglycosylation events become more favorable in the presence of a higher concentration of short chains and neutral pH [12]. Consequently, the overall degradation is an inherently complex process, making a global kinetic analysis a challenging task.

Recently, we have investigated the degradation of HA catalyzed by Bovine testicular Hyal (BTH) in the presence of PEG 6000 crowders using a CE-based enzymatic assay [13]. BTH specifically cleaves the β 1,4-glucosaminidic bond between C1 of the glucosamine moiety and C4 of the glucuronic acid [14] and the reaction products (oligomers) thus contain $2m$ monomers ($m$ =1,2,3 …). The analysis of reaction mixtures with CE allowed us to evaluate the speciation of oligosaccharides during the late stage of the reaction. Our data revealed a pronounced, non-trivial inhibition of the reaction in the presence of PEG 6 k (w/w). In order to identify and assess quantitatively the key reaction steps during the second stage of HA hydrolysis, a reliable kinetic model is required to fit the experimental speciation observed under different conditions. The aim of the present paper is to formulate such a model and parameterize it against accurate experimental data. More precisely, we present the theoretical background of the model and its application to the kinetics of HA degradation in the absence of crowding and in the presence of 10 % and 17 % PEG 6 k (w/w). The global analysis of these complex systems allows us to determine the relevant rate constants, elucidate the main reaction mechanism and pathways and, more generally, get considerable insight into the effects of crowding on the functioning of enzymatic reactions within biological environments.

## 2. Results

In our previous work [13], the kinetics of BTH was followed at two HA concentrations (0.3 and 1 mg/mL). In addition, for each HA concentration, the reaction mixture contained different quantity of the PEG crowder (0, 10, 17 % w/w). For all these experiments, each performed in duplicates, the relative composition of oligomers within the mixture was determined after electrophoretic separation by CE [13]. Here we develop a global kinetic model and expound the rational analysis of these original data.

### *2.1. Two-phase modeling approach*

The early stages of the fragmentation kinetics of HA appear exceedingly complex. Moreover, the time resolution of our kinetic measurements did not allow us to monitor accurately enough the evolution of long intermediate oligomers, the longest fragments whose concentration could be determined accurately enough being 10/12-mers. However, our method allowed us to gather reliable information about the time evolution of the populations of shorter oligomers (2-, 4-, 6-, 8- and 10-mers) for late stages, depending on the HA concentration and crowding conditions.

Our kinetic scheme is illustrated in Fig. 1 (see Methods for the detailed mathematical formulation). We consider here that, starting from some time $t_0$ (our time origin that defines the onset of the *late* phase), the total number of monomers $M_0$ corresponding to the initial concentration of HA is entirely contained in oligomers up to dodecamers. In the subsequent kinetic transformations, the total mass is progressively fed to the population subspace comprising $2m$-ers with $m = 1,2,3,4,5$ (see Fig. 1), until the final stationary state is reached.

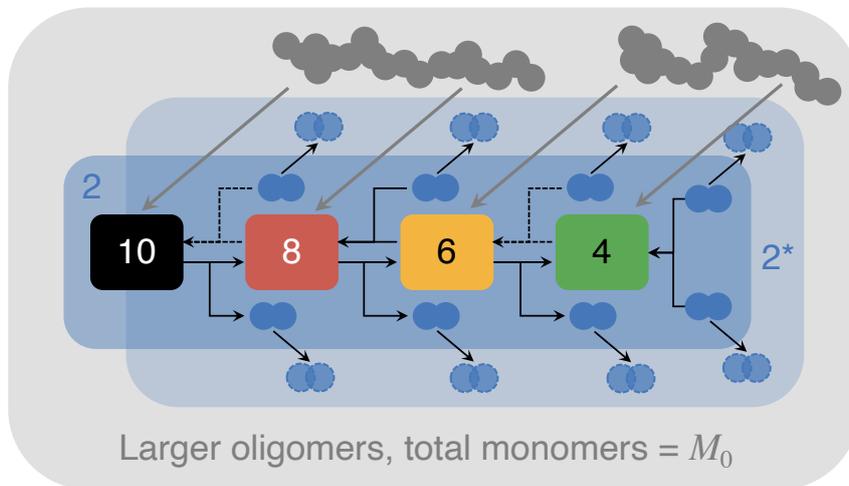

**Figure 1**. Kinetic model of the *late* stage of HA degradation introduced in this paper. In the *early* stage of HA degradation, large oligomers successively undergo enzymatic hydrolysis until the total number of monomers $M_0$ is entirely contained in oligomers up to 12-mers (light grey box). Starting from time $t_0$ (our time origin), such reservoir starts feeding the subspace of oligomers up to 10-mers, where successive hydrolysis and transglycosylation events occur until the stationary state is attained. Pathways implying fragmentation or recombination reactions not comprising at least one dimer molecule were found to be absent, as well as some comprising dimers (dashed arrows). The asymptotic state of the kinetics in the late-stage subspace proceeds until the total mass $M_0$ is fed entirely into tetramers and (likely enzyme-bound) inactive dimer species, noted as 2* (see Methods). We stress that the choice of 10-mers as the largest oligomers in the *late* stage is somewhat arbitrary (our theoretical

scheme is general), and is dictated only by the resolution of our CE-based kinetic measurements.

The late phase comprises both *fragmentation* (hydrolysis) and *recombination* (transglycosylation) events. We denote with $K_f$ the fragmentation rate for the hydrolysis of the parent oligomer $2m$ ($m$ = 2-5) to any oligomer $2(m-1)$ (e.g. 8 → 2 + 6) and $K_g$ the transglycosylation rate for the recombination reactions (*e.g.* 2 + 6 → 8). Moreover, we found that, in order to reproduce the experimental data, one should account for a process of (nearly) irreversible inactivation of the dimer species in the active site of BTH. These dimers are allowed to be transformed into inert (*i.e.* no more available to be incorporated into longer oligomers) species through addition of a water molecule [6] that cannot be enzymatically stitched back to other oligomers in a recombination event. It is worthy to note that using CE hyphenated to high-resolution mass spectrometry we were able to detect dimers in the reaction media [7].

Fig. 2 illustrates how the total mass at the onset of the late degradation stage of HA depends on both (initial) HA concentration and PEG crowding. It can be seen that when the total substrate mass was increased from 0.3 mg/mL to 1 mg/mL, the total mass at the onset of the late stage was estimated self-consistently to be larger by the same factor as expected (inset in Fig. 2). This confirms the soundness of our approach. Furthermore, Fig. 2 illustrates a first intriguing result of our analysis. For a given concentration of HA, the total mass at the onset of the late stage appears to decrease with increasing crowding concentration (Fig. 2, main panel). As the total mass of HA is the same for a given concentration, this suggests that crowding makes an increasing portion of the substrate virtually inaccessible to the enzyme, likely due to steric hindrance effects. At the higher concentration of crowding considered (17 % w/w), it appears that between 15 and 20 % less of substrate has been degraded at the end of the early stage. Quite reasonably, this effect is slightly less pronounced for the higher concentration of substrate.

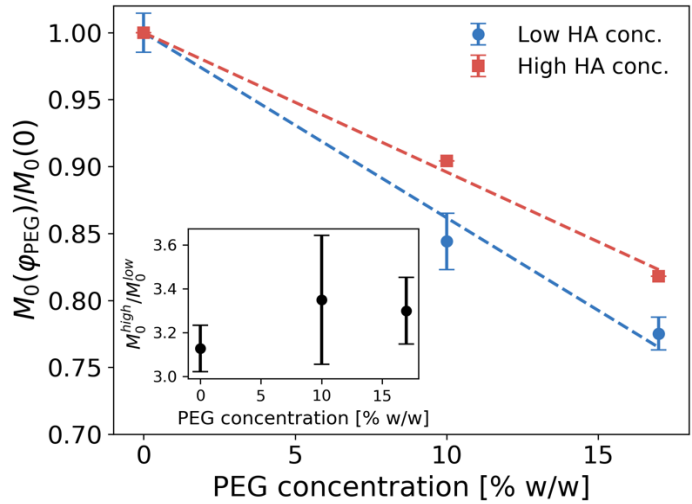

**Figure 2.** Total mass at the onset of the late degradation stage of HA degradation *vs* crowding, normalized to the value obtained in the absence of PEG. Inset: illustration of the effect of HA concentration on its enzymatic degradation in the early phase: increasing HA concentration by a factor of 3 (from 0.3 mg/mL to 1 mg/mL) leaves a total mass about 3 times as large at the onset of the late stage (see also Methods).

### *2.2. Global fit of the degradation kinetics*
Fig. 3 shows the results of global fits of our model to the experimentally determined molar fractions of oligomers in the late-stage subspace ($m$ < 6) for the 6 parameter choices comprising 2 HA concentrations (0.3 and 1 mg/mL) and 3 different crowding conditions, *i.e.* 0, 10 % and 17 % PEG (see Methods for the details). We found a very good agreement for all 6 experiments despite the limited number of points available at 1 mg/mL of HA.

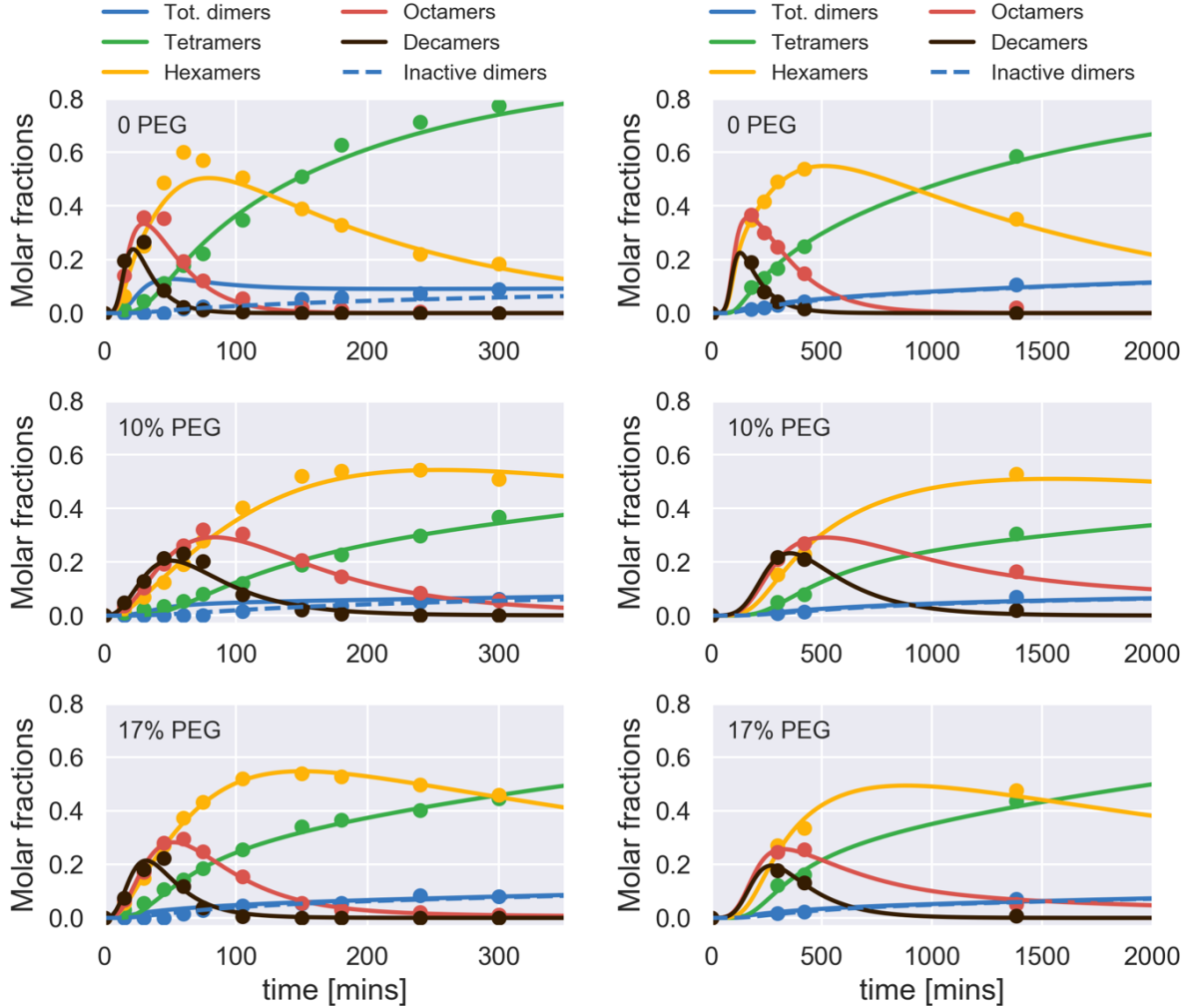

**Figure 3**. Time evolution of species containing up to 10 monomers, without crowding and in the presence of different concentrations of PEG (10 and 17%), for low HA concentration, 0.3 g/L (left panels), and high HA concentration, 1 mg/mL (right panels). Experiments (symbols) and global fit with the kinetic model (6). The blue dashed lines represent the molar fraction of inert dimers, $f_2^*(t)$, while the solid blue lines represent the total molar fractions of dimers, $f_2(t) + f_2^*(t)$, i.e. enzyme-bound plus inactive.

The fitted rates for all different conditions are summarized in Table I (low HA concentration) and Table II (high HA concentration). Interestingly, although all possible reactions are in principle taken into account in the model, it immediately appeared from the fitting procedure that several pathways are virtually absent in the mechanism that leads to the asymptotic state (*i.e.* rate → 0, see also Methods).

**Table I.** Low HA concentration, 0.3 g/L. Best-fit values of the fragmentation and transglycosylation rates, inactivation rate of dimers, injection weights $w_m$ and asymptotic fractions of dimeric species and tetramers predicted by our model, averaged over different independent runs of the stochastic optimization algorithm. Uncertainties correspond to statistical errors on the mean. Rates are in units of minutes$^{-1}$.

| Kinetic parameter | PEG Concentration (w/w) | | |
|---|---|---|---|
| | 0 | 10 % | 17 % |
| 10 → 8 + 2 | 0.0699 ± 0.0003 | 0.0304 ± 0.0005 | 0.0581 ± 0.0005 |
| 10 → 6 + 4 | 0 | 0 | 0 |
| 8 → 6 + 2 | 0.0410 ± 0.0001 | 0.0160 ± 0.0003 | 0.0288 ± 0.0004 |
| 8 → 4 + 4 | 0 | 0 | 0 |
| 6 → 4 + 2 | 0.005683 ± 0.00003 | 0.001160 ± 0.000005 | 0.001886 ± 0.000005 |
| 2 + 2 → 4 | 0.301 ± 0.001 | 4 ± 3 | 19.8 ± 0.1 |
| 4 + 2 → 6 | 0 | 0 | 0 |
| 6 + 2 → 8 | 0 | 0.07 ± 0.032 | 0.27 ± 0.03 |
| 6 + 4 → 10 | 0 | 0 | 0 |
| 8 + 2 → 10 | 0 | 0 | 0 |
| 2 → 2* | 0.00339 ± 0.00004 | 0.014 ± 0.005 | 0.0416 ± 0.0002 |
| $w_3$ | 0.179 ± 0.001 | 0.182 ± 0.003 | 0.195 ± 0.002 |
| $w_4$ | 0.332 ± 0.003 | 0.258 ± 0.006 | 0.177 ± 0.006 |
| $w_5$ | 0.488 ± 0.003 | 0.560 ± 0.007 | 0.628 ± 0.007 |
| $f_2^{*\infty}$ | 0.09488 ± 0.00002 | 0.1651 ± 0.0002 | 0.16732 ± 0.00005 |
| $f_4^{\infty}$ | 0.90511 ± 0.00001 | 0.8348 ± 0.0002 | 0.83254 ± 0.00005 |

**Table II.** High HA concentration, 1 g/L. Best-fit values of the fragmentation and transglycosylation rates, inactivation rate of dimers, injection weights $w_m$ and asymptotic fractions of dimeric species and tetramers predicted by our model, averaged over 20 different independent runs of the stochastic optimization algorithm. Uncertainties correspond to statistical errors on the mean. Rates are in units of minutes$^{-1}$.

| Kinetic parameter | PEG Concentration (w/w) | | |
|---|---|---|---|
| | 0 | 10 % | 17 % |
| 10 → 8 + 2 | 0.01063 ± 0.0001 | 0.00456 ± 0.00002 | 0.00630 ± 0.00009 |
| 10 → 6 + 4 | 0 | 0 | 0 |
| 8 → 6 + 2 | 0.00585 ± 0.00002 | 0.0113 ± 0.0004 | 0.0152 ± 0.0004 |
| 8 → 4 + 4 | 0 | 0 | 0 |
| 6 → 4 + 2 | 0.000711 ± 0.000001 | 0.000156 ± 0.000002 | 0.000350 ± 0.000002 |
| 2 + 2 → 4 | 12 ± 3 | 18 ± 2 | 18 ± 2 |
| 4 + 2 → 6 | 0 | 0 | 0 |
| 6 + 2 → 8 | 0 | 2.8 ± 0.2 | 2.0 ± 0.2 |
| 6 + 4 → 10 | 0 | 0 | 0 |
| 8 + 2 → 10 | 0 | 0 | 0 |
| 2 → 2* | 0.018 ± 0.002 | 0.0136 ± 0.0007 | 0.014 ± 0.001 |
| $w_3$ | 0.240 ± 0.002 | 0.071 ± 0.003 | 0.006 ± 0.005 |
| $w_4$ | 0.366 ± 0.004 | 0.418 ± 0.004 | 0.56 ± 0.01 |
| $w_5$ | 0.394 ± 0.004 | 0.511 ± 0.005 | 0.439 ± 0.009 |
| $f_2^{*\infty}$ | 0.16231 ± 0.00006 | 0.1905 ± 0.0006 | 0.1492 ± 0.0005 |
| $f_4^{\infty}$ | 0.83769 ± 0.00001 | 0.8095 ± 0.0006 | 0.8508 ± 0.0005 |

| $\tau_M$ [min] | 97.6 ± 0.9 | 267.8 ± 0.6 | 209.3 ± 0.7 |

As a remarkable general finding, the kinetic model represented in Fig. 1 was found to explain all experiments performed with the same accuracy, no matter the substrate concentration and the crowding conditions. This means that changing the total number of monomers in the system (HA concentration) and/or reducing the available space through crowding agents does not lead to major changes in the physics of the late stage of HA degradation. Within the same kinetic universality class (identified by the scheme in Fig. 1), the overall main conclusion is that increasing HA concentration and decreasing available space through crowding (i.e. increasing effective reactant concentrations) are both factors whose individual effect is to slow down HA degradation. Furthermore, when acting simultaneously (e.g. high HA concentration and 10 % PEG), their effects seem to combine in a simple fashion (see again Fig. 2).

The degradation process of HA yields tetramers and dimers as final products. The asymptotic fractions $f_n^\infty$ ($n = 2,4$) estimated within our model are reported in Tables I and II. Our analysis confirms that tetramers are indeed not hydrolyzed [9] and appear only as products of the slow hydrolysis of hexamers, the shortest substrate of BTH, or as fast recombination of dimers. Interestingly, we found that the rate of other cleavage and recombination steps involving tetramers did not affect appreciably the kinetics (in more technical terms, those parameters were systematically pushed to the lower null bound in the least-square minimization). This observation excludes such pathways from the overall mechanism of the late-stage degradation. On the contrary, dimers are systematically cleaved away from longer oligomers $2m$ ($m > 2$). Quite generally, these *active* dimers appear to be available for recombination only over a transient period of the kinetics and correspond to the dimer bound in the enzyme active site [8]. A fraction of them decays in an inert form that can no longer be used in catalyzed recombination events and accumulates in the mixture. This inactivation is formally compatible with the transfer of an "active" dimer to water as acceptor suggested by Kakizaki. [8]

Our analysis reveals that there exists one main recombination reaction, namely $2 + 2 \rightarrow 4$. The corresponding rate, $K_g(2, 2| 4)$, turns out to be the fastest of all recombination rates. On the contrary, the reaction $6 + 2 \rightarrow 8$ became significant only in the presence of crowding, suggesting that such relatively unfavorable recombination requires some confinement to become thermodynamically favorable. Moreover, the rate of the reaction $8 + 2 \rightarrow 10$ tends to zero even in the presence of PEG. This points out that recombination events involving longer oligomers are intrinsically strongly limited, despite experimental evidence for such transglycosylation reported by several authors [8, 9, 12]. We found that recombination rates are generically much faster than hydrolysis rates (see again Table I and Table II), at least by one order of magnitude. However, the global rate-limiting step is clearly the hydrolysis of hexamers.

At high HA concentration, dimer inactivation and recombination pathways are significantly faster with respect to lower HA concentration and the asymptotic fraction of inactive dimers almost double (~ 9 % → 16 %, Tables I and II). On the contrary, the rate $K_f$ (6|4,2) as well as those of similar fragmentation steps, decreases ~ 7-8 times when HA concentration increases from 0.3 to 1 mg/mL of HA. This global slowdown is consistent with the formation of an electrostatic non-productive complex HA-BTH, as suggested in [10], which causes an effective decrease of the concentration of free enzymes.

### 2.3. *Effect of crowding on reaction rates*
In general, it can be affirmed that crowding has the clear effect of (i) boosting recombination events and (ii) slowing down the rate-limiting hexamer hydrolysis as well as faster fragmentation pathways (see again Tables I and II). Accordingly, crowding also causes a substantial slowdown of the kinetics. Most likely, this effect originates in excluded-volume (confinement/caging) effects that raise the effective concentration of oligomers around dimer-loaded enzymes. However, interestingly, this effect turns out to be non-monotonic, as PEG appears to slow down the kinetics more at a weight fraction of 10 % than it does at 17 %. This is visible in the kinetics of the total mass build-up shown in Fig. 2, as well as in the kinetics of individual molar fractions displayed in Fig. 3. However, in our previous work using CE-based hyaluronidase assay [13], we stressed out that the high viscosity of the PEG solution at 17 % can hinder the introduction of the reaction solution into the capillary. Therefore, the results can be slightly underestimated at high crowding.

The predicted asymptotic fraction of tetramers decreases appreciably in the presence of crowding, from about 90 % to about 83 %. Correspondingly, the asymptotic fraction of accumulated inert dimers asymptotically nearly doubles from about 10 % to about 17 % in the presence of crowding (see Tables I and II). At the same time, the transient increase of the "active" dimers that can be used in recombination events is also substantially quenched in the presence of crowding and become almost negligible. The probability that a dimer be transferred to water instead of another oligomer increases in the presence of PEG, which is reflected by a significant increase of inactivation rates at low HA concentrations. Interestingly, one may surmise that there exists a concentration-dependent threshold for such inactivation, as the corresponding rate, $k^*$, no longer increases with crowding when the substrate is present at higher (enough) concentration (see Table II). This results in similar asymptotic fractions of inactive dimers (~15- 19 %), regardless of whether PEG crowders are present at all and at what concentration.

## 3. Conclusions and perspectives

In this work, we have developed a full kinetic model for the late stage of HA degradation catalyzed by BTH. Our theory was applied to rationalize kinetic experiments through the analysis of samples by capillary electrophoresis at two different concentrations of HA, in the absence and at different concentrations of PEG crowders, in order to investigate how excluded volume affects enzymatic digestion of HA.

Remarkably, the overall degradation scheme that emerges from our analysis is universal, regardless of HA concentration as well as of degree of crowding, and agrees with several previous observations made separately on some of the kinetic steps that we included in our comprehensive scheme.

The rates obtained by fitting simultaneously the time-dependence of all the oligomers in the late-stage subspace (dimers to decamers) led us to detailed quantitative insight into the enzymatic hydrolysis of HA. More precisely, the enzymatic hydrolysis in the late phase is dominated by the transfer of the "active" dimer, whereby BTH works in an exolytic mode [11]. Interestingly, we find clear confirmation that water molecules may act as acceptors, leading to the accumulation of silenced (inert) dimers that can no longer be stitched back to other oligomers. Overall, we demonstrate that, depending on HA concentration and crowding fraction, the asymptotic state of HA degradation comprises between 10 % and 20 % of (inactive) dimers and between 80 % and 90 % of tetramers.

In conclusion, we have shown that faster recombination steps are favoured by crowding coming from both PEG and HA itself (self-crowding), while hydrolytic steps are strongly hindered by those factors. Beyond the quantitative insight that the present theory affords, our work clearly illustrates the biological importance of hetero- and self-crowding in the regulation of enzymatic activity. Consequently, crowding appears as a tool that can be used in mechanistic studies to help elucidate the key reaction steps of biochemical transformations. It should be remarked that, provided the concentration of higher oligomers ($m > 5$) can be determined accurately enough, the present model can be easily extended to include chemical transformations comprising longer oligomers.

Finally, we note that it would be of interest to apply the model presented in this paper to data collected at higher pH. As this would favor transglycosylation, one could test whether the kinetics would still be described within the universality class of the present model. Furthermore, it would be interesting to develop a detailed microscopic model describing the non-monotonic inhibition of the cleavage of HA in the presence of increasing concentrations of PEG. To this regard, this would entail considering the effect of other kinds of crowding agents, such as dextran and other more compact polysaccharides (e.g. Ficoll), to explore to what extent excluded-volume interactions are sensitive to the conformation and chemical properties (e.g. hydrophobicity) of the space-filling species.

## 4. Methods

### 4.1. Experimental procedure.

The degradation of HA (0.3 and 1 mg/ml) by the BTH was investigated in the presence of 0, 10 and 17 % of PEG 6000 (w/w). A sample of the reaction mixture was taken at different incubations times and its composition was analyzed using capillary electrophoresis (CE). The corrected peak area (CPA) of electropherogram peaks was used as a reliable mean for the quantification of different oligomers. Further experimental details were described in [13].

### 4.2. Kinetic model.

*General considerations.*

Let us imagine to set the time origin at time $t_0$ such that the total number of monomers $M_0$ (obviously conserved) is entirely contained in oligomers longer than a certain size $2n_{max}$. Let us indicate with $C_{2m}$ the concentration of the species containing $2m$ monomers for $1 \leq m \leq n_{max}$ and with $M(t)$ the total concentration of the oligomers longer than $2n_{max}$. A reasonable way to describe the kinetics from this moment onwards is as follows:

$$\begin{cases} \dfrac{dC_{2m}}{dt} = F_m(\vec{C}) + \left(\dfrac{k_m}{2m}\right) M \\ \dfrac{dM}{dt} = -\left(\sum_{m=1}^{n_{max}} k_m\right) M \end{cases} \quad (1)$$

with initial conditions $M(0) = M_0$, $\vec{C}(0) = 0$. The terms $F_m(\vec{C})$, which in principle are each a function of all the concentrations $\vec{C} \equiv \{C_2, C_4, \ldots, C_{2n_{max}}\}$, represent the fragmentation and re-association kinetics in the same subspace of state space. Moreover, the rate equations for the concentrations $C_{2m}$ have source terms that represent the further fragmentation of larger species contained in the macro-species $M(t)$. Equations (1) are valid under the hypothesis that the species $\vec{C}$ only transform among themselves and do not associate to yield oligomers larger than $2n_{max}$. Mathematically, this translates to the condition

$$2 \sum_{m=1}^{n_{max}} m\, F_m(\vec{C}) = 0 \quad (2)$$

As a consequence, Equations (1) conserve the total number of monomers, that is,

$$M(t) + 2 \sum_{m=1}^{n_{max}} m F_m(\vec{C}) = M_0 \qquad (3)$$

where $M_0$ is the total number of monomers present in the system and is fixed by the initial concentration of hyaluronic acid. In principle, the macro-species $M$ can be eliminated by using Eq. (3), so that the kinetics in the state subspace spanned by the concentrations $\vec{C}$ can be written as

$$\frac{dC_{2m}}{dt} = F_m(\vec{C}) + \frac{k_m}{2m}\left(M_0 - 2\sum_{m=1}^{n_{max}} m\, C_{2m}\right) \qquad (4)$$

and solved with the initial condition $\vec{C} = 0$.

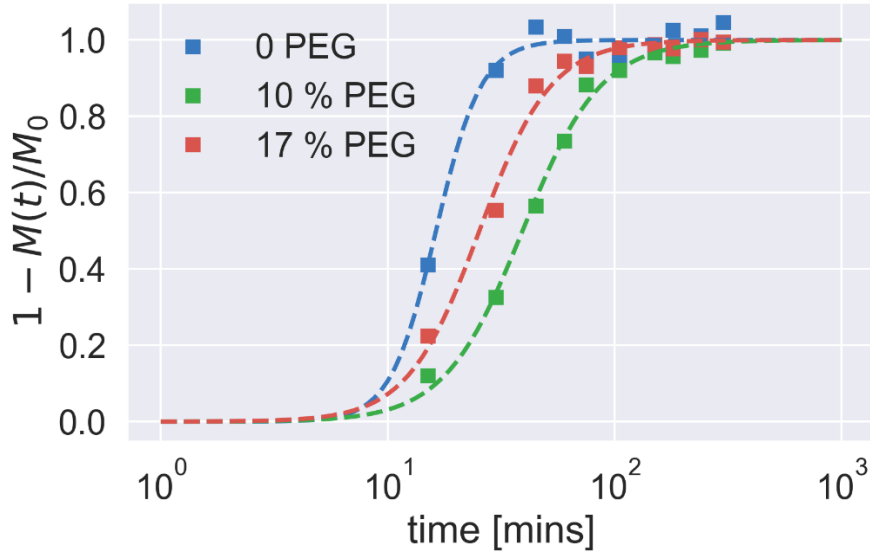

**Figure 4.** Low HA concentration, 0.3 g/L. Time evolution of the total normalized mass in the subspace spanned by the species $\{C_2, C_4, ..., C_{10}\}$, $2\sum_{m=1}^{5} mC_{2m}(t) = M_0 - M(t)$, as measured experimentally (symbols), setting the origin of time at the onset of the late phase (appearance of decamers). The dashed lines are plots of fits performed with the function (5). Best- fit parameters are $(M_0, \tau_M, \beta)$ = (5072.9, 16.315, 4.3409) for 0 PEG, (4282.5, 39.259, 2.5121) for 10 % PEG and (3933.1, 25.341, 2.7444) for 17 % PEG ($M_0$ in CPA, time in minutes).

The above model predicts that the concentration of the macro-species $M$ should vanish exponentially while its monomer content is injected in the state subspace of $\vec{C}$. However, the experimental data are far better fitted with a Hill function of the kind (see for instance Fig. 2 for low HA concentration 0.3 mg/ml)

$$M_{\text{obs}} = \frac{M_0}{1 + (t/\tau_M)^\beta} \tag{5}$$

To be more precise, one expects the true consumption kinetics of *M(t)* to be a weighted sum of exponentials, as this macro-species represents indeed the weighted superposition of many oligomers of unknown length and with unknown proportions. As it is often the case, time decay of many (typically more than 3) exponential signals with weights not differing by many orders of magnitude can be well approximated by a power-law decay. In practice, based on this observation, we replaced the kinetics (4) in the $(1, n_{\text{max}})$ subspace with the following equations

$$\begin{cases} \dfrac{dC_{2m}}{dt} = F_m(\vec{C}) - \dfrac{w_m}{2m} \dfrac{dM_{\text{obs}}}{dt} \\ \dfrac{dM}{dt} = \dfrac{dM_{\text{obs}}}{dt} \end{cases} \tag{6}$$

with initial condition $M(0) = M_0$, $\vec{C}(0) = 0$. The role played by the different injection rates $k_m$ in Eqs. (4) is played in Eqs. (6) by the set of normalized weights $w_m$, with $\sum_{m=1}^{n_{\text{max}}} w_m = 1$. Eqs. (6) conserves the total mass, $2\sum_{m=1}^{n_{\text{max}}} mC_{2m}(t) + M(t)$, and *force* injection of mass from the reservoir *M(t)* into the shorter oligomers $\{C_2, C_4, \ldots, C_{2n_{\text{max}}}\}$ according to the temporal law observed experimentally.

### 4.3. Fragmentation model in the subspace $\{C_2, C_4, \ldots, C_{2n_{max}}\}$.

Based on our experimental observations for low HA concentration, we fix $n_{\text{max}} = 5$, that is, we consider species from dimers up to decamers. Species $\vec{C}$ can be enzymatically fragmented and recombined. Accordingly, the most general form of the kinetics terms reads

$$F_m(\vec{C}) = \sum_{n=m+1}^{n_{\text{max}}} K_f(2n|2(n-m), 2m) C_{2n}$$

$$- \sum_{m_1} \sum_{m_2=m-m_1} K_f(2m|2m_1, 2m_2) C_{2m}$$

$$+ \sum_{m_1} \sum_{m_2=m-m_1} K_g(2m_1, 2m_2|2m) C_{2m_1} C_{2m_2} \tag{7}$$

$$- \sum_{n=1}^{n_{\text{max}}-m} K_g(2m, 2n|2(m+n)) C_{2m} C_{2n}$$

Here we have indicated with $K_f$ and $K_g$ the fragmentation and transglycosylation rates (probabilities per unit time), respectively, such that:

- $K_f(2m|2m_1, 2m_2)$ is the probability *per* unit time that a 2*m*-mer be split into one 2*m*₁-mer and one 2*m*₂-mer.

- $K_g(2m_1, 2m_2|2m)$ is the probability *per* unit time that a bond is formed between a $2m_1$-mer and a $2m_2$-mer to form a $2m$-mer.

In order to reproduce the experimental data, the dimers are allowed to be transformed into inert species that cannot be enzymatically stitched back to other oligomers. Let us indicate with $C_2^*$ the concentration of inert dimers. With the addition of this reaction equations (6) become

$$\begin{cases} \dfrac{dC_{2m}}{dt} = F_m(\vec{C}) - \dfrac{w_m}{2m}\dfrac{dM_{\text{obs}}}{dt} & \text{for } 1 \leq m \leq 5 \\ \dfrac{dC_2}{dt} = F_1(\vec{C}) - k^*C_2 - \dfrac{w_1}{2}\dfrac{dM_{\text{obs}}}{dt} \\ \dfrac{dC_2^*}{dt} = k^*C_2 \\ \dfrac{dM}{dt} = \dfrac{dM_{\text{obs}}}{dt} \end{cases} \quad (8)$$

with initial conditions $M(0) = M_0$, $\vec{C}(0) = 0$, $C_2^*(0) = 0$. In order to fit model (8) to the experimental data, we used a stochastic minimization routine based on Storn and Price original algorithm, [15] to minimize the following global cost function

$$\chi^2 = 4 \sum_{i=1}^{N_p(1)} [C_2^*(t_i) - C_2^{\text{exp}}(t_i)]^2 + \sum_{m=2}^{n_{\max}} \sum_{i=1}^{N_p(m)} m^2 [C_{2m}(t_i) - C_{2m}^{\text{exp}}(t_i)]^2 \quad (9)$$

with respect to the set of rates $K_f$, $K_g$, $k^*$ and the set of weights $\vec{w}$. Here $N_p(m)$ indicates the number of time points of the time series for the concentration of $2m$-mers. Furthermore, we found that better fits were obtained by including the concentration of inactive dimers only in equations (9). This is confirmed by the best-fit values of the rates of recombination events of the kind 2–mers + 2n–mers → 2(n + 1)-mers, which turned out to be at least one order of magnitude faster than the fragmentation rates. Furthermore, we found that the best fits were obtained by including injection channels from the macro-species $M$ to decamers, octamers and hexamers only, these in turn fragmenting further to generate the shorter oligomers. This amounts to choosing the following set of weights $\vec{w} \equiv (w_1, w_2, \ldots, w_5)$

$$\vec{w} = (0, 0, w_3, w_4, w_5), \quad \text{with} \quad w_3 + w_4 + w_5 = 1 \quad (10)$$

The asymptotic values of the molar fraction of the species that constitute the final product of the overall fragmentation of hyaluronan are defined as $f_{2m}^\infty \equiv \lim_{t \to \infty} f_{2m}(t)$, where the molar fractions read

$$\begin{cases} f_{2m}(t) = \dfrac{mC_{2m}(t)}{\sum_{n=1}^{n_{\max}} mC_{2m}(t) + C_2^*(t)} \\ f_2^*(t) = \dfrac{C_2^*(t)}{\sum_{n=1}^{n_{\max}} C_{2m}(t) + C_2^*(t)} \end{cases} \quad (11)$$

In all cases considered, it is apparent that the final products are tetramers and inert dimers.

### *4.4. Fragmentation kinetics at high HA concentration.*

The kinetics at high HA concentration (1 g/L) were sampled at a lower rate than those at lower concentration. Therefore, the total mass recorded, including oligomers up to 16mers, saturates quickly, making an independent fit of the global mass build-up such as those shown in Fig. 1 impossible. However, we found that the model (eq 8) still allowed us to reproduce the experiments to an excellent extent. In practice, in order not to multiply the fitting parameters, we have found that the best strategy was to fix the Hill exponent β. The best-fit values of the fitting parameters are reported in tables I and II and leave the global time constant $\tau_M$ free to float alongside the rates $K_f$, $K_g$ the injection weights $\vec{w}$ and the dimer inactivation rate $k^*$. We have found that the values of β required to reproduce the observed build-up of species up to decamers are higher than at lower substrate concentration. This is expected due to the presence of longer oligomers that contribute additional (exponential) injection channels in the 2-10-mer populations. In practice, we have fixed β = 6 in the absence of crowding, and β = 4 in the presence of 10 % and 17 % of PEG, which is the same trend as observed in the fits of the total mass build-up at low substrate concentration shown in Fig. 4.

**Author Contributions**

RNe, JH and FP planned enzymatic experiments; RNa and LO performed experiments; FP and JH analyzed the data and wrote the paper. All authors discussed the results and contributed to the final manuscript.

**Acknowledgements**
This work has been supported by the University of Orléans (France), the CNRS (Centre National de la Recherche Scientifique, France), the Federation of research FR2708 and the Labex SynOrg (ANR-11-LABX-0029).

**References**

1. Vercruysse, K. P., Lauwers, A. R. & Demeester, J. M. (1995) Kinetic investigation of the action of hyaluronidase on hyaluronan using the Morgan-Elson and neocuproine assays, *Biochemical Journal.* **310**, 55-59.
2. Vercruysse, K. P., Lauwers, A. R. & Demeester, J. M. (1995) Absolute and empirical determination of the enzymatic activity and kinetic investigation of the action of hyaluronidase on hyaluronan using viscosimetry, *Biochem J.* **306 ( Pt 1)**, 153-160.
3. Fang, S., Hays Putnam, A. M. & LaBarre, M. J. (2015) Kinetic investigation of recombinant human hyaluronidase PH20 on hyaluronic acid, *Anal Biochem.* **480**, 74-81.
4. Reissig, J. L., Storminger, J. L. & Leloir, L. F. (1955) A modified colorimetric method for the estimation of N-acetylamino sugars, *The Journal of biological chemistry.* **217**, 959-66.

5. Asteriou, T., Deschrevel, B., Delpech, B., Bertrand, P., Bultelle, F., Merai, C. & Vincent, J.-C. (2001) An Improved Assay for the N-Acetyl-d-glucosamine Reducing Ends of Polysaccharides in the Presence of Proteins, *Analytical Biochemistry.* **293**, 53-59.
6. Vercruysse, K. P., Lauwers, A. R. & Demeester, J. M. (1994) Kinetic investigation of the degradation of hyaluronan by hyaluronidase using gel permeation chromatography, *Journal of chromatography B, Biomedical applications.* **656**, 179-90.
7. Fayad, S., Nehmé, R., Langmajerová, M., Ayela, B., Colas, C., Maunit, B., Jacquinet, J.-C., Vibert, A., Lopin-Bon, C., Zdeněk, G. & Morin, P. (2017) Hyaluronidase reaction kinetics evaluated by capillary electrophoresis with UV and high-resolution mass spectrometry (HRMS) detection, *Analytica Chimica Acta.* **951**, 140-150.
8. Kakizaki, I., Ibori, N., Kojima, K., Yamaguchi, M. & Endo, M. (2010) Mechanism for the hydrolysis of hyaluronan oligosaccharides by bovine testicular hyaluronidase, *The FEBS Journal.* **277**, 1776-1786.
9. Hofinger, E. S. A., Bernhardt, G. & Buschauer, A. (2007) Kinetics of Hyal-1 and PH-20 hyaluronidases: Comparison of minimal substrates and analysis of the transglycosylation reaction, *Glycobiology.* **17**, 963-971.
10. Lenormand, H., Amar-Bacoup, F. & Vincent, J.-C. (2013) Reaction–complexation coupling between an enzyme and its polyelectrolytic substrate: Determination of the dissociation constant of the hyaluronidase–hyaluronan complex from the hyaluronidase substrate-dependence, *Biophysical Chemistry.* **175-176**, 63-70.
11. Deschrevel, B., Tranchepain, F. & Vincent, J.-C. (2008) Chain-length dependence of the kinetics of the hyaluronan hydrolysis catalyzed by bovine testicular hyaluronidase, *Matrix Biology.* **27**, 475-486.
12. Saitoh, H., Takagaki, K., Majima, M., Nakamura, T., Matsuki, A., Kasai, M., Narita, H. & Endo, M. (1995) Enzymic Reconstruction of Glycosaminoglycan Oligosaccharide Chains Using the Transglycosylation Reaction of Bovine Testicular Hyaluronidase, *Journal of Biological Chemistry.* **270**, 3741-3747.
13. Nasreddine, R., Orlic, L., Banni, G. A. H. D., Fayad, S., Marchal, A., Piazza, F., Lopin-Bon, C., Hamacek, J. & Nehmé, R. (2020) Polyethylene glycol crowding effect on hyaluronidase activity monitored by capillary electrophoresis, *Analytical and Bioanalytical Chemistry.* **412**, 4195-4207.
14. Lee, A., Grummer, S. E., Kriegel, D. & Marmur, E. (2010) Hyaluronidase, *Dermatologic Surgery.* **36**, 1071-1077.
15. Storn, R. & Price, K. (1997) Differential Evolution – A Simple and Efficient Heuristic for Global Optimization over Continuous Spaces, *J of Global Optimization.* **11**, 341–359.